\documentstyle[11pt, epsf]{article} 
\normalbaselines

\newcommand{\be}{\begin{equation}}
\newcommand{\ee}{\end{equation}}

\newcommand{\lr}{\left|}
\newcommand{\rr}{\right\rangle}

\begin{document}

\begin{center}
{\Large {\bf Quantum entanglement in the voltage dependent sodium channel can reproduce the salient features 
of neuronal action potential initiation}}

\vspace{3mm}

Johann Summhammer$^{a,}$\footnote{Email address: summhammer@ati.ac.at} 
and Gustav Bernroider$^{b,}$\footnote{Email address: Gustav.Bernroider@sbg.ac.at}
\end{center}

\vspace{3mm}

$^a$ Vienna University of Technology, Atominstitut, Vienna, Austria

$^b$ Dept. of Organismic Biology, Neurosignaling Unit, University of Salzburg, Austria

\vspace{5mm}

{\bf Abstract}

{\small 
We investigate the effects of a quantum entanglement regime within an ion conducting molecule (ion channel) of the neuronal 
plasma membrane on the onset dynamics of propagating nerve pulses (action potentials). In particular, we model the onset 
parameters of the sodium current in the Hodgkin Huxley equation as three similar but independent probabilistic mechanisms 
which become quantum entangled. The underlying physics is general and can involve entanglement between various degrees of 
freedom underlaying ion transition states or 'gating states' during conduction, e.g. Na$^+$ ions in different channel locations, 
or different 'affinity' states of ions with atoms lining the sub-regions of the channel protein ('filter-states').We find that the 
'quantum corrected' Hodgkin Huxley equation incorporating entangled systems states can reproduce action potential pulses with the 
critical onset dynamics observed recently in neocortical neurons in vivo by Naundorf et al. [Nature {\bf 440}, 1060 (20 April 2006)]. 
Interestingly,  the suggested entanglement term can also slow down action potential initiation.
}


\vspace{5mm}

\section{Introduction}

Since its introduction in 1952 the Hodgkin Huxley equation \cite{HH} (HH-equation) has been very successful in describing
neuronal pulses \cite{Bialek} \cite{Koch} and the behavior of networks of neurons (e.g. \cite{Sentao Wang}). 
The conceptual basis of the HH-model equation is given by the stochastic motion of ions 
through channel proteins within the cell membrane, driven by potential differences across the membrane and regulated by potential 
dependent conductances. The formalization of the corresponding model of classical electrodynamics, in which the
membrane lipid bilayers play the role of a capacitor, leads directly to the HH-equation.

However, a recent experiment on cortical neurons has shown a fast onset of the action potential that can not be reproduced by 
the classical HH-equation \cite{Naundorf}. The authors recover the observed pulse shapes only by postulating a correlation
between the neighbouring ion conducting channels of the cell membrane. However, such cooperations among channels beyond the constraining transmembrane field is at variance with the dominating fluid-mosaic model of the plasma membrane \cite{Singer} and the independent 'gating' mechanisms intrinsic to the HH model. 

The present paper approaches this problem from a microscopic point of view. We identify possible quantum effects within a 
{\em single} ion conducting channel which can lead to different rates of the transfer of ions through the channel and hence
to a deviation of the shape of the voltage pulse behind the classical HH-equation. The main motivation for a quantum mechanical approach to ion conduction mechanisms stems from the small dimensions and short range forces among and between ions and the surrounding atomic structure of the ion hosting protein. The atomic details and the functionality of the K$^+$ channel (\cite{MacKinnon1}, 
\cite{MacKinnon2}, \cite{Berneche}, \cite{Luzhkov}) and of the Na$^+$ channel \cite{Shi} have been revealed during the last years.
These findings suggest that both, the potassium ions as well as the sodium ions have to pass through a narrow pore 
in a one-by-one queue or 'hopping' fashion. Because the diameter of the available passage is on the order of a nanometer and there 
are potential minima along the axis of passage that can temporarily 'trap' an ion within a few angstroms (e.g. \cite{Berneche}),
the passage of an ion may be more similar to a quantum transition from one three-dimensional trap to the next
\cite{BernroiderRoy}, rather than a classical random motion under electrostatic guiding forces. 

Within this frame, one can think of at least three different types of quantum effects which could play a role. 
First, there can be a one-particle quantum effect in which the passage of an ion is described by a wavefunction rather 
than by a classical motion of a point particle. Along this view the channel structure would function as a position and time
dependent potential to which the particle is exposed. With the known dimensions, an ion's quantized energy levels within 
a temporary trap along the channel will exhibit a spacing of 0.1\% to 1\% of $kT$ at 300 K. The conductance through the channel 
would then no longer follow Ohm's law as implicitly assumed in the HH-equation, but would exhibit quantized steps as 
observed with thin metallic wires, also at room temperature \cite{GarciaNanoWires}. 

A second effect involves the interaction of successive ions in the channel. In this case ions occupying specific channel locations
at a given moment (e.g.two to three ions in the 'filter region', one ion in the 'cavity region' as in the KcsA potassium channel)
are described by one common and entangled wavefunction. The channel structure and intervening water molecules are treated as in case one, by a position and time dependent potential to which the ions are exposed. 
As discussed earlier in \cite{BernroiderRoy}, the coordinated arrangement of ions in the channel \cite{Shi},\cite{MacKinnon2} 
is conspicuously similar to designed linear ion traps that are currently investigated as possible candidates for quantum computers 
\cite{Walther.IonTraps}, \cite{Duan.IonTraps}, \cite{Lewenstein.IonTraps}. 
In this case the entanglement is ensured by the repulsion between ions. Overall, it can be expected that this situation leads to a well coordinated passage of ions through the channel. Probabilities of ion-state transitions will appear highly correlated, leading to conductances that can be expected to be different from the classical diffusion model underlaying the HH-equation. 

A third possibility is offered by quantum effects due to the 
atomic structure of the channel itself. Here one considers the ions as classical particles, but treats the
channel structure as a quantum system whose various atoms are described by one common wave function. This view
is particularly supported by MD simulation studies on ion-oxygen coordination indicating that an ion passing the channel 'experiences'
a sequence of four 'oxygen-cages' each provided by four in-plane carbonyl oxygen atoms (e.g. see \cite{Guidoni}). 
The binding of these oxygen atoms to their parent backbone amino acid structure and their typical energy of $kT$ can be expected to force them into some quantized vibrational state. 
The interaction term between these atoms can evolve into a common entangled quantum state. As a consequence, the motion of each 
oxygen atom will depend on the motion of all other oxygen atoms. Since this motion, which quantum mechanically results 
from a superposition of at least two energy states of the oxygen atom, has a considerable amplitude \cite{Berneche},
it will in turn influence the probability with which an ion may pass through a plane of oxygen atoms. The ion would then experience
the channel as a series of correlated gates: After one in-plane passage, the ion may have a higher chance to 
pass the next one, etc. Again, this situation implies consequences in ion conductances which can be expected to deviate from the classical picture.

In real situations all of these effects may play a role and there additional factors may arise, depending on how much of the molecular structure
of the channel is described at the atomic scale. In the present paper 
we will focus on the second and third aspects, which involve entanglement. However, we do not need to choose any of the described scenarios specifically, because both aspects share similar effects in the view of a transition from classical probabilities to quantum probabilities. We can  remain fairly general and speak of several 
independent degrees of freedom which will become quantum entangled. These degrees of freedom can be given by the 
motion of several independent ions, or be provided by the interacting oxygen atoms. In any case, they can also be given by 
some other probabilistic variables which are relevant for the transmission of charge through the channel.
Yet another reason for a fairly general treatment lies in the classical understanding of the conduction
mechanism behind the HH-equation. Here one assumes the existence of 'gates', whose exact nature is not specified, but
which control the magnitude of the ion current. Besides their overall voltage dependence, these 'gates' are considered to 
behave in a probabilistic way. This aspect, together with the mentioned atomic resolution scale where the gating variables must be instantiated, naturally calls for a quantum theoretical description.
Here we describe the 'gates' of the Na$^+$ channel quantum mechanically, because the Na$^+$ ions are responsible
for the initial rise of the pulse in the HH-equation. We expected that quantum entanglement might shed some light on the observed 'faster-than-expected' action potential initiation (API) mechanisms as found by \cite{Naundorf}. Indeed, we find that entanglement can lead to a faster onset of the pulse. We also find that the classical pulse is recovered when decoherence destroys the introduced entanglement term.

The paper is organized as follows. Section 2 reviews the classical Hodgkin Huxley equation. Section 3 presents the quantum mechanical
model for the transit of ions in the Na$^+$ channel and introduces quantitative considerations on the necessary coherence in the thermal 
environment. Section 4 presents the results of a parametric case study. Section 5 discusses the results and suggests possible experimental 
consequences.

\section{The classical Hodgkin Huxley equation}

\noindent
The Hodgkin-Huxley equation describing the action potential discharge of a piece of space clamped axon
(e.g. \cite{Bialek}, \cite{Sentao Wang}) is given by 
\be
C\frac{dV}{dt} = I_{ext} - g_{K}n^4 (V-E_K) - g_{Na}m^3h(V-E_{Na}) - g_L (V-E_L).
\ee
Here, $g_K$, $g_{Na}$, and $g_L$ are the conductances through the membrane for $K^+$, $Na^+$ and leakages of
all other ions, respectively. $E_K$, $E_{Na}$ and $E_L$ are the corresponding equilibrium potentials. The
capacitance of the patch of membrane is given by $C$ and the membrane potential by $V$. $I_{ext}$ denotes an 
external current, which can arise from a pulse from a neighbouring piece of membrane. The variables
$n$, $m$, and $h$ are the gating variables. These are voltage and time dependent according to,
\be
\frac{dn}{dt} = \alpha_n(V)(1-n) - \beta_n(V)n,
\ee
\be
\frac{dm}{dt} = \alpha_m(V)(1-m) - \beta_m(V)m,
\ee
\be
\frac{dh}{dt} = \alpha_h(V)(1-h) - \beta_h(V)h.
\ee
The gating variables can assume numerical values between 0 and 1 and represent the probabilities that 
the specific conditions for the conduction of ions through a membrane channel are met at a given time 
and transmembrane voltage. The variables $\alpha_n$,$\alpha_m$,$\alpha_h$ and $\beta_n$,$\beta_m$,$\beta_h$, exhibit exponential 
dependencies on voltage, which ultimately result from probabilistic rate considerations in the molecular 
picture of the stochastic motion of ions. Figure 1 demonstrates a typical pulse produced by eq.(1).

\begin{figure}[ht]
\begin{center}
\epsffile{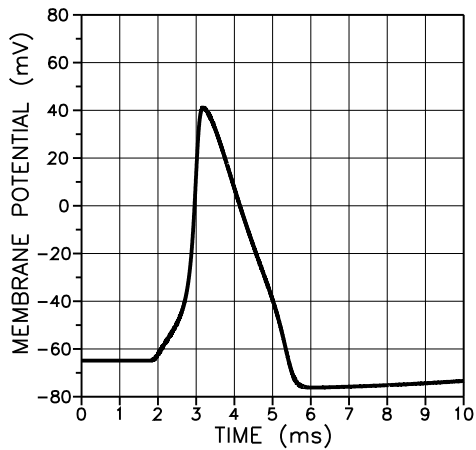}
\caption[Fig.1:]{A typical voltage pulse of a patch of space clamped neuronal membrane as simulated with the classical 
Hodgkin Huxley equation. Parameters:
$C=1\mu F/cm^2$, $E_K=50mV$, $E_{Na}=-77mV$, $E_L=-54mV$, $g_K=36mS/cm^2$, $g_{Na}=120mS/cm^2$, $g_L=0.3mS/cm^2$, and
(in units of $1/ms$ with voltages in $mV$) 
$\alpha_n(V)=0.01(V+55)/(1-e^{-(V+55)/10})$, $\beta_n(V)=0.125e^{-(V+65)/80}$,
$\alpha_m(V)=0.1(V+40)/(1-e^{-(V+40)/10})$, $\beta_m(V)=4e^{-(V+65)/18}$,
$\alpha_h(V)=0.07e^{-(V+65)/20}$ and $\beta_h(V)=1/(1+e^{-(V+35)/10})$. The size of the membrane patch was assumed as
60 $\mu m \times$ 60 $\mu m$. The pulse was triggered by an external current pulse of trapezoidal shape. It started to rise
linearly at 1.8 $ms$, reached 1 $nA$ at 2.0 $ms$, remained constant until 3.0 $ms$ and then dropped linearly to reach zero
at 3.2 $ms$. } 
\end{center}
\end{figure}

\section{Inserting quantum entanglement into the HH-equation}

There is an accumulating list of evidence emerging within the new frame of atomic scale resolution studies of ion conducting proteins that point to the necessity to involve quantum effects in the description of transmembrane ion permeation. The path of an ion through the channel is at most a few nanometers (the 'filter' region of KcsA potassium channels extends to 1,2 nm). With a thermal velocity of $v\approx\sqrt{3kT/m}$ such an ion moves 
somewhat less than a nanometer per picosecond (0.570 nm/ps for sodium ions, as estimated from their observed diffusion coefficients in electrolyte solutions \cite{Kuyucak}). If the quantum entanglement between several degrees of freedom has a 
coherence time of just a few picoseconds, this will be longer than the average passage time through the filter region of the
channel and quantum mechanical deviations from classical behavior will not be washed out. 
Within the conception where ions passing the channel constitute the entangled quantum system and the molecular structure 
of the channel acts as the environment, we find basically two options. Coherence times may or may not be long enough to allow a few ions passing as
a single quantum system. The factors which can destroy this possible entanglement are given by the fast random motions of this structure. The ions are in electrostatic contact with the oxygen atoms of the carbonyl groups and with the intervening water molecules. These
atoms and molecules are exposed to thermal energies so that their motional degrees of freedom, expressed either as vibrational modes in the case of the oxygen atoms or as translational motion in the case of the water molecules, have velocities in the order of 10$^2$ to 10$^3$ m/s, similar to the
velocity of free ions. Therefore, ions in transit will not experience fast changes in the channel structure. For example an ion might
'experience' about 10 vibrational periods of any of the oxygen atoms, while it remains within a ion-oxygen distance of a few angstroms. This
will influence the relative phases of the eigenstates of the quantum state of the ions in a way so that the overall properties of their
entangled state will be influenced.  However, it is unlikely that the ion will be subjected to a sudden disruption, and therefore the evolution through the channel can  be coherent. 
The measurement of the current will involve the translocation of a very large number of ions (up to $10^8$ ions/s).
Although the transit of groups of ions could possibly occur as entangled quantum states, successive cohorts, or ions in different channels,
will undergo different quantum evolutions due to the thermal dynamics of the molecular structure of the channel. The measured current
will therefore be an average over different quantum evolutions. We may expect this to constitute the main source of decoherence under the realistic many particle situation.

The situation appears to be similar when the oxygen atoms of the carbonyl groups are treated as quantum systems and the ions as classical systems.
Successive ions states within a single channel and of course ions hosted by different channels will be confronted with different entangled quantum motions of the coordinating oxygen atoms. This will at some times enhance and some other times impede the transition of ion states in contrast to a classical state transition. Again, because only the average is measurable, the ensemble will be subjected to decoherence. 
In both models, however, we cannot anticipate necessarily that the averaging over different quantum evolutions will ultimately eliminate the quantum effects. This supports the involvement of a quantum analysis and allocates a role for decoherence. Within the present calculations this role will be modeled by the introduction of a tunable parameter reflecting a varying degree of entanglement between the engaged subsystems of the process. 

We set out by noting that in eq.(1) sodium as well as potassium channel dynamics are controlled by the products
of gating variables, reflecting probabilities. For potassium this is  the $n^4$ term and for sodium this is the
$m^3h$ term. Within the framework of classical stochastic processes we can interprete these 'gates' as representing a mechanism in which 
several {\em independent} conditions must be met to allow for peak ion conduction to occur.
Here we are particularly interested in the $m^3(V)$ activation curve, because it is responsible for 
the rate of membrane depolarization that shapes the action potential initiation segment of the neuronal pulse.
In the following we will treat $m$ quantum mechanically and leave $h$, the term that expresses the fraction of channels available for activiation, as a classical probability.
The fact that $m$ is raised to the third power 
suggests that three similar but independent conditions must be met to facilitate peak sodium conductance 
\cite{Bialek}. In the present quantum mechanical context this corresponds to three independent degrees of freedom. Obviously, the molecular 
structure and the number of ions which are in transit within the channel at any given moment would indicate some more degrees of freedom.  
We will however confine the present investigation to the frame set by the classical model. Furthermore, we assume that each degree of freedom can only reflect two different states, corresponding to {\em open} and {\em closed}, respectively.

In order to integrate quantum concepts into the HH-equation we construct the following scenario: within a given channel, at a given voltage of the membrane and at a given time a certain quantum state composed from the three degrees of freedom will have evolved. After a typical coherence time
$\tau$, which will be in the order of picoseconds following the arguments above, this quantum 
state will collapse due to coupling to the surrounding thermal environment. Current will only flow, if 
the collective state function  collapses to the state where all three degrees of freedom are in a condition facilitating conduction. Within the view of several ions in the channel treated as one quantum system, we may assume that one or more ions can pass the filter region. In the model where
the oxygen atoms of the cage structure are treated as the quantum system, it is implied that one or more of the ions previously caged within the filter region can pass. If the collective state collapses to any one of the other possibilities (e.g., only two of the three conditions are met), no current will flow. After the collapse, a new quantum state will evolve where the amplitudes for the various possibilities will depend on the present membrane voltage and moment of time during the emergence of the action potential. This state will again collapse after a typical decoherence time $\tau$, and these sequences will be repeated during the entire duration of the pulse.
 
Let us now look more closely at the classical factor $m^3$. According to eq.(3) $m$ depends on voltage and time. Thus at a
given voltage and time there is a probability of $m^3$ that all three conditions required for conduction are met and current can flow.
But there is also a probability that only one condition is true and two are false, given by $3m(1-m)^2$, or that two conditions are
true and one is false, given by $3m^2(1-m)$, or that all three conditions are false, given by $(1-m)^3$. Since none of these
situations permits the flow of current, they don't appear in the HH-equation. However, quantum mechanically we have to take them 
into account, because the quantum state of the three degrees of freedom is always a superposition of all possibilities.
For the sake of simplicity we may assume that the three degrees of freedom have bosonic character, so that it is only important how many of them are in the {\em open} or in the {\em closed} condition, but not which ones. The underlaying state space is then four dimensional with the basis 
vectors $\lr 0\rr$, $\lr 1\rr$, $\lr 2\rr$, and $\lr 3\rr$. The numbers denote how many of the 
degrees of freedom are in the {\em open} condition. If we want to write the classical situation in quantum mechanical
form, leaving aside some arbitrary phases, we arrive at the state
\be
\lr\psi\rr_{cl} =  \sqrt{(1-m)^3}\lr0\rr + \sqrt{3m(1-m)^2}\lr1\rr + \sqrt{3m^2(1-m)}\lr2\rr +\sqrt{m^3}\lr3\rr.
\ee
The size of all amplitudes representing this state depends only on one parameter, that is $m$. However, the general
quantum state of the three degrees of freedom is:
\be
\lr\psi\rr =  a_0\lr0\rr + a_1\lr1\rr + a_2\lr2\rr +a_3\lr3\rr,
\ee
where $a_0$, ..., $a_3$ are complex amplitudes with no restrictions, except the normalization condition 
$\sum_{j=0}^3|a_j|^2 =1$. Neglecting an overall phase of the state we now have {\em six} free parameters.
In order to reduce this number in our study, and because it is of little consequence for the general
result, we can assume that the states $\lr1\rr$ and $\lr2\rr$ have the same amplitude $\epsilon$, thus
\be
\epsilon \equiv a_1 = a_2
\ee
This simplification implies that at any moment it is equally likely to find either one degree of freedom or two degrees of freedom in
the {\em open} condition. 
In the next step we make the connection to the classical probability $m$ by setting $a_0$ proportional to $\sqrt{1-m^3}$ and $a_3$
proportional to $\sqrt{m^3}$.  Again we make a simplification and use the same proportionality constant $\delta$ for both relations.
This step assumes equality in the quantum mechanical deviation from the classical case for all three parameters to be in the {\em open} 
conditions with all three parameters to be in the {\em closed} condition. Now the quantum state becomes
\be
\lr\psi\rr = \delta \sqrt{1-m^3}\lr0\rr + \epsilon \lr1\rr + \epsilon \lr2\rr + \delta \sqrt{m^3}\lr3\rr.
\ee
We can set $\delta$ as real, since the overall phase of the quantum state is irrelevant. The 
normalization requirement $\lr\langle\psi|\psi\rangle\right|^2=1$ leads to the relation
between $\epsilon$ and $\delta$,
\be
2\lr\epsilon\right|^2 = \left[1 - \delta^2 \left(1-3m+3m^2\right)\right],
\ee
so that $\delta^2$ is the only free parameter. Because of $2\lr\epsilon\right|^2 \ge 0$ and $2\lr\epsilon\right|^2 \le 1$
(from normalization) the possible range of values for $\delta^2$ is
\be
0\le|\delta^2|\le \frac{1}{1-3m+3m^2}.
\ee
Within this range we can distinguish between $\delta^2<1$ and $\delta^2>1$. The probability that the channel is open at a given moment
is given by the probability that $\lr\psi\rr$ will collapse to $\lr3\rr$, which is $\delta^2 m^3$. Therefore, quantum states with 
$\delta^2=1$ behave just like classical states. Quantum states with $\delta^2<1$ lead to a reduction of Na$^+$ conduction and 
quantum states with $\delta^2>1$ lead to an enhancement of Na$^+$ conduction.
Without a detailed atomic model it cannot be stated whether the total state space available for $\lr\psi\rr$ encompasses more possibilities for the
reduction or for the enhancement of ion conduction. However, the classical case $\delta^2=1$ is just {\em one} possibility and thermodynamic
averaging over {\em all} quantum possibilities does not necessarily converge into the classical case. The average could be tilted in one or the other direction 
and consequently traces of the quantum behavior will appear.

Now we obtain the quantum mechanical correction for the original HH-equation by replacing the term $m^3$ by the
term $\delta^2m^3$. Note however, that according to eq.(10) $\delta^2$ can assume a range of values for a given $m$. This will depend on
the particular quantum state prevalent at a given moment. Clearly, the quantum state of the three degrees of freedom will change
every time it is recovered following a collpase, depending on the environmental interaction. Therefore, the effective 
value of $\delta^2$ is likely to vary stochastically within its range during a neuronal pulse. We capture this
situation by replacing the parameter $\delta^2$ by another parameter $\kappa$, which expresses the degree of entanglement. We normalize this magnitude so that it can assume values between -1 and 1. The case $\kappa=0$ will represent the classical situation of $\delta^2=1$, with no entanglement. 
The case $\kappa=1$ will represent maximum positive entanglement, with $\delta^2 = (1-3m+3m^2)^{-1}$. For positive values of $\kappa$
we thus have a deviation from the classical HH-equation which results in an increased conduction rate of Na$^+$ ions, described by the quantum-corrected 
HH-equation:
\be
C\frac{dV}{dt} = I_{ext} - g_{K}n^4 (V-E_K) - g_{Na}\left[1+\kappa\frac{3m(1-m)}{1-3m(1-m)}\right] m^3h(V-E_{Na}) - g_L (V-E_L).
\ee
In the case of negative entanglement the three degrees of freedom tend to be not just independent but de-synchronized, such
that if one of them is in the {\em closed} condition the other two are in the {\em open} condition, and vice versa. This impedes
the transport of Na$^+$ ions and is expressed by values of $\delta^2$ between 0 and 1. We rewrite this in terms of $\kappa$, which we now define as
$\kappa \equiv \delta^2 -1$ and obtain the corrected HH-equation for negative entanglement
\be
C\frac{dV}{dt} = I_{ext} - g_{K}n^4 (V-E_K) - g_{Na}\left[1+\kappa\right] m^3h(V-E_{Na}) - g_L (V-E_L).
\ee

\section{Results}

The behavior of neuronal pulses was studied as a function of $\kappa$, a measure of 'm-gating entanglement' in the sodium current activitation dynamics of the Hodgkin-Huxley model. Equations (11) and (12) as well as equations (2), (3) and (4) were
integrated numerically by rewriting them as difference equations and adding up the time steps.  
The voltage pulse was initiated by a trapezoidal current pulse as defined in the caption of Fig.1. Typical results are demonstrated in Fig.2. 
\begin{figure}[ht]
\begin{center}
\epsffile{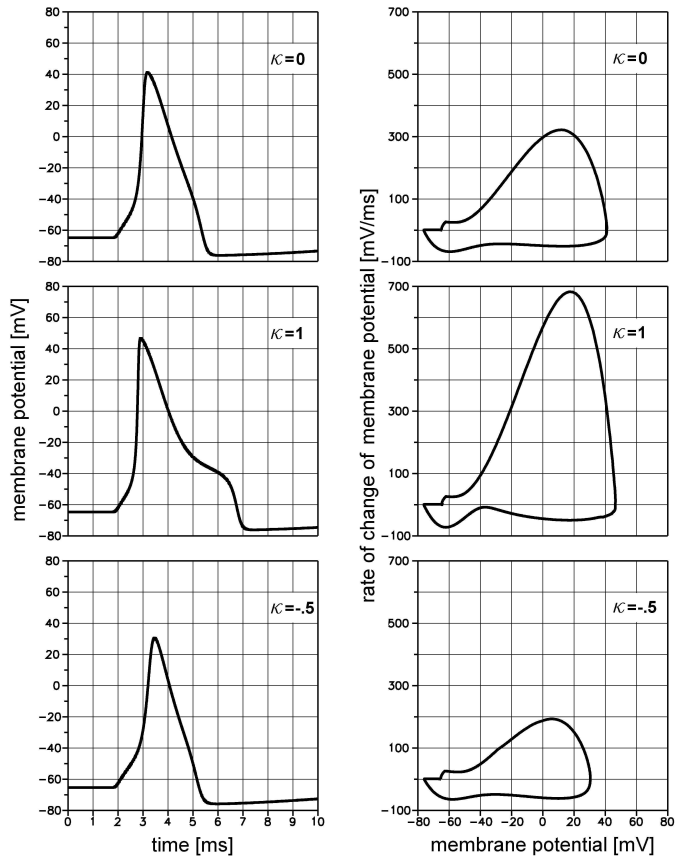}
\caption[Fig.2:]{ Voltage pulse (left) and phase plot (right) for different values of $\kappa$ ( other parameters set as indicated in Fig.1). The
top inserts show the classical case ($\kappa=0$), the middle row demonstrates the case of maximum positive quantum entanglement,
($\kappa=1$) and the bottom row gives an example of negative quantum entanglement ($\kappa=-.5$). } 
\end{center}
\end{figure}
The top pictures show the classical situation ($\kappa=0$). The effects of positive quantum entanglement is demonstrated in the middle inserts for
$\kappa=1$. In this signal the action potential pulse exhibits a sharp inital rise of the membrane voltage with a higher velocity as apparent from the phase plot. The signals slope $dV/dt$ within the rising part becomes more than twice as large as in the classical case. The maximum is reached 
1.1 $ms$ after the beginning of the injected current, whereas in the classical case this onset takes 1.4 $ms$. Thus the quantum-corrected m-gating dynamics is characterized by a clear increase in the onset-rapidness of the action potential initiation dynamics. Interestingly, positive quantum entanglement 
also influences the decaying part of the voltage pulse. The small shoulder in the decaying part of the signal, which is already visible in the classical pulse, becomes accentuated with increasing values of $\kappa$. To interprete this behaviour one has to take attention to the part of the voltage pulse that is most sensitive to the introduced quantum entanglement term. 
According to eq.(10) this can be expected to occur at those times when the gating probability $m$ reaches the value of $1/2$. In the model of three independent classical
'gates' this implies a probability of 1/8 that all three of them are open simultaneously. But at that point the quantum mechanical enhancement factor
$\delta^2$ reaches a value of 4, so that in the model of quantum entanglement of three degrees of freedom the effective probability for 
{\em open} is 1/2. This is responsible both for a faster rise
during the onset of the pulse and for a slower decay in the falling phase of the pulse. However, in the present model the inactiviation dynamics of sodium channels as expressed by the 'gating variable' 'h' 
is kept classical. A QM interpretation of this term could well be expected to compensate the 'shoulder effect' observed in the present 'hybrid' 
semi-classical treatment. 

For negative quantum entanglement one can observe the opposite effect. This is shown for $\kappa=-0.5$ in the bottom row of Fig.2. 
The rising part of the pulse becomes slower, and the falling part becomes faster as can clearly be seen from the phase plot. Altogether this leads to a lower voltage peak. Investigation of the onset of the pulse under different injected currents also indicates, that the stimulus puls threshold must be higher under the negative entanglement regime as compared with the classical case or the case of positive quantum entanglement. 
At a maximum negative entanglement of $\kappa=-1$ only a sub-threshold behaviour could be observed.

\section{Summary and Discussion}

We have analyzed how quantum entanglement of ions or atoms within a Na$^+$ conducting ion channel can lead to an observable
effect on the shape of a neuronal action potential. In particular we argue that entangled states can evolve repeatedly. Each time the state exists only for
a coherence time in the order of picoseconds, this being sufficient to exert coherent control over the passage of a few ions through the channel. Within
the full duration of a transmembrane voltage pulse, which scales along milliseconds, a succession of some 10$^8$ to 10$^9$ different entangled quantum states can form and collapse within each channel molecule. Some of these states will lead to a conductance behavior of ions that is not different from classical conduction, while other states 
will exhibit highly quantum entangled and non-intuitive transition probabilities of ions. The measurement of a voltage pulse will thus show an average over an ensemble of different quantum states. These quantum states are characterized by various amplitudes and phases. Because only one 
component of a given quantum state is assumed to be 'permissive' allowing the conduction of ions, we combine these amplitudes to just one parameter which accounts for the average degree of entanglement.  

We find that quantum entanglement can significantly enhance the transit probabilities of ions through a channel. The experimental observation reported
by Naundorf et al \cite{Naundorf} about signatures of real action potential initiation that can not be reproduced by the classical, HH-type equation of motion, seem to be compatible with the present predictions made from a quantum physical approach. In particular, the characteristic features behind signal onset-rapidness and rate of potential changes in dependence of membrane voltage are predicted by the present model. This reproduces essential aspects of the observed signal dynamics without the necessity to call for unestablished inter-channel cooperativity (e.g see the discussion of McCormic et al, \cite{McCormick})
We also find the possibility that quantum entanglement reduces the transit probability of ions through a channel.Insight into the relation between conductance enhancing and reducing effects of quantum physical interpretations will depend on very detailed atomic models and a complete quantitative description of
coupling strengths between and within the involved atoms and ions. This is beyond the present possibilities.

With regard to an experimental assessment of the proposed hypothesis 
on quantum entangled ion-molecule states we can expect that the variation of the spontaneous quantum states
of either ions or carbonyl oxygen atoms, or both, will become smaller at lower temperatures. At these lower temperatures we can expect that the influence of quantum entanglement on the voltage pulse dynamics will become more pronounced. Another experimental indication would be to compare fermionic and bosonic Na isotopes with respect to their behaviour during ion conduction. Naturally, $^{23}$Na is the only isotope and its ion Na$^+$ is a fermion. On the other hand, the ion Na$^+$ of the artificial isotope $^{22}$Na,
which has a half life of 2.6 years, is a boson. It may therefore be easier to achieve entanglement with bosonic characteristics, as we have 
assumed in our model calculations, with $^{22}$Na$^+$ as compared with $^{23}$Na$^+$. But one should note however, that bosonic characteristics are not required 
in principle to exert an influence of multi-ion quantum entanglement on conduction mechanisms. In the present work we have introduced the bosonic character for reasons of simplification and approximation. We expect, that an experimental comparison of neuronal pulses with all sodium ions provided by $^{22}$Na with neuronal pulses made from
$^{23}$Na isotopes could still give valuable information whether quantum entanglement between the Na$^+$ ions in the narrow pore of an ion channel 
plays a role in conduction as suggested here. In theory similar considerations apply to the role of vibrational modes of the oxygen atoms forming the filter structure through which the ions must pass. Here both $^{16}$O (abundance 99.762\%)
and $^{17}$O (abundance 0.038\%) are natural and stable. The former is a boson and the latter a fermion. 

Here we have paid attention to quantum entanglement in Na$^+$ conduction. Similar theoretical models could be developed for K$^+$ conduction.
For potassium there are also two natural isotopes available. One is a fermion ($^{40}$K, abundance 0.012\%, half life 1.277$\times$10$^9$ years) and one 
is a boson ($^{39}$K, abundance 93.26\%, stable), with reversed characteristics of the corresponding single charged ions. A comparison of experiments with 
$^{39}$K$^+$ to experiments with $^{40}$K$^+$ could therefore help to elucidate possible quantum entanglement during ion conduction within the highly conserved K$^+$ channel family. 

In the present approach we have not explicitly addressed the newly emerging delicate interplay between ion selectivity, permeation rates and different levels of gating states (e.g. affinity filter states correlating with molecular 'constriction gates' and blocked filter states correlating with activated and inactivated channel states), as observed in different voltage sensitive channels (eg \cite{Berneche2005} \cite{VanDongen}). The emerging picture uncovers correlations of coordinated single ion states with mechanisms controlling the access of ions to permeation, instantiated at different structural scales of the proteins. Many of the suggested processes seem to be good candidates behind the 'formal agents' expressed as quantum mechanical degrees of freedom in the present theoretical frame.


\begin{thebibliography}{99}

\bibitem{HH} A.L. Hodgkin and A.F. Huxley, J. Physiol., {\bf 463}, 391 (1952); 
{\em A quantitative description of membrane current and its application to conduction and excitation in nerve}

\bibitem{Bialek} Blaise Ag\"uera y Arcas, Adrienne L. Fairhall, and William Bialek, Neural Computation {\bf 15}, 1715 (2003);
{\em Computation in a Single Neuron: Hodgkin and Huxley Revisited}

\bibitem{Koch} Christof Koch and Ojvind Bernander, in [The Handbook of Brain Theory and Neural Networks\", eds. Michael A. Arbib],The MIT Press, 129 (1995); {\em Axonal Modeling}

\bibitem{Sentao Wang} Sentao Wang, Feng Liu, and Wei Wang, Phys.Rev.E{\bf 69}, 011909 (2004), 
{\em Impact of spatially correlated noise on neuronal firing} 

\bibitem{Naundorf} Bj\"orn Naundorf, Fred Wolf and Maxim Volgushev, Nature {\bf 440}, 1060 (20 April 2006);
{\em Unique features of action potential initiation in cortical neurons}

\bibitem{Singer} S.J.Singer, G.L. Nicholson, Science {\bf 175}, 720 (1972);
{\em The fluid mosaic model of the structure of cell membranes}

\bibitem{MacKinnon1} Joao H. Morals-Cabral, Yufeng Zhou and Roderick MacKinnon, Nature {\bf 414}, 37 (1 November 2001);
{\em Energetic optimization of ion conduction rate by K$^+$ selectivity filter}

\bibitem{MacKinnon2} Yufeng Zhou, Joao H. Morals-Cabral, Amelia Kaufman and Roderick MacKinnon, Nature {\bf 414}, 43 (1 November 2001);
{\em Chemistry of ion coordination and hydration revealed by a K$^+$ channel-Fab complex at 2.0 \AA resolution}

\bibitem{Berneche} Simon Bern\`eche and Benoit Roux, Nature {\bf 414}, 73 (1 November 2001);
{\em Energetics of ion conduction through the K$^+$ channel}
 
\bibitem{Luzhkov} Victor B. Luzhkov, Johan $\AA$qvist, Biochemica et Biophysica Acta {\bf 1548}, 194 (2001);
{\em K$^+$/Na$^+$ selectivity of the KcsA potassium channel from microscopic free energy perturbation calculations}

\bibitem{Shi} Ning Shi, Sheng Ye, Amer Alam, Liping Chen and Youxing Jiang, Nature {\bf 440}, 570 (23 March 2006);
{\em Atomic structure of a Na$^+$- and K$^+$-conducting channel} 

\bibitem{BernroiderRoy} Gustav Bernroider and Sisir Roy in [Fluctuations and Noise in Biological, Biophysical and Biomedical Systems III\",
eds. Nigel Stocks, Derek Abbott, Robert P. Morse], Proc. of SPIE Vol 5841, 205 (2005); 
{\em Quantum entanglement of K$^+$ ions, multiple channel states and the role of noise in the brain}
  
\bibitem{GarciaNanoWires} J. L. Costa-Kr\"amer, N. García, P. García-Mochales, P. A. Serena, M. I. Marqués, and A. Correia, 
Phys.Rev.B {\bf 55}, 5416 (1997); 
{\em Conductance quantization in nanowires formed between micro and macroscopic metallic electrodes} 

\bibitem{Walther.IonTraps} 
Jiannis Pachos and Herbert Walther, Phys. Rev. Lett. {\bf89}, 187903 (2002); 
{\em Quantum Computation with Trapped Ions in an Optical Cavity}  

\bibitem{Duan.IonTraps} 
Shi-Liang Zhu, C. Monroe, and L.-M. Duan 
Phys. Rev. Lett. {\bf 97}, 050505 (2006); 
{\em Trapped Ion Quantum Computation with Transverse Phonon Modes}
 
\bibitem{Lewenstein.IonTraps}
Marisa Pons, Veronica Ahufinger, Christof Wunderlich, Anna Sanpera, Sibylle Braungardt, Aditi Sen(De), Ujjwal Sen, and Maciej Lewenstein 
Phys. Rev. Lett. {\bf 98}, 023003 (2007); 
{\em Trapped Ion Chain as a Neural Network: Error Resistant Quantum Computation}

\bibitem{Guidoni}
L. Guidoni and P. Carloni, Biochimica et Biophysica Acta {\bf 1563}, 1 (2002);
{\em Potassium permeation through the KcsA channel: a density functional study}

\bibitem{Kuyucak}
S. Kuyucak, O. S. Andersen and Shin-Ho Chung, Rep. Prog. Phys. {\bf 64}, 1427 (2001); 
{\em Models of permeation in ion channels}

\bibitem{McCormick}
David A McCormick, Y Shu and Yuguo Yu, Nature, {\bf 445}, E1 (2007);
{\em Hodgkin and Huxley model - still standing ?}

\bibitem{Berneche2005}
S. Berneche and B. Roux, Structure, {\bf13}, 592 (2005);
{\em A gate in the selectivity filter of potassium channels}

\bibitem{VanDongen}
A.M.J. Van Dongen, PNAS, {\bf 101}, 8644 (2003);
{\em K-channel gating by an affinity-switching selectivity filter}

 
\end{thebibliography}
\end{document}